\documentclass[sigconf]{acmart}
\renewcommand\footnotetextcopyrightpermission[1]{} % removes footnote with conference information in first column
\def\BibTeX{{\rm B\kern-.05em{\sc i\kern-.025em b}\kern-.08emT\kern-.1667em\lower.7ex\hbox{E}\kern-.125emX}}

\copyrightyear{2018}
\acmYear{2018}
\setcopyright{acmlicensed}
\acmConference[Woodstock '18]{Woodstock '18: ACM Symposium on Neural Gaze Detection}{June 03--05, 2018}{Woodstock, NY}
\acmBooktitle{Woodstock '18: ACM Symposium on Neural Gaze Detection, June 03--05, 2018, Woodstock, NY}
\acmPrice{15.00}
\acmDOI{10.1145/1122445.1122456}
\acmISBN{978-1-4503-9999-9/18/06}

\usepackage{bm}
\usepackage{subcaption} % for sub figure
\usepackage{tikz} % for latex plot
\usetikzlibrary{decorations.pathreplacing}
\usetikzlibrary{backgrounds, scopes, fit}
\usetikzlibrary{shapes.arrows, shapes.geometric}
\usetikzlibrary{datavisualization}
\usetikzlibrary{datavisualization.formats.functions}
\usepackage[ruled,vlined]{algorithm2e}

\newcommand{\specialcell}[2][c]{%
  \begin{tabular}[#1]{@{}c@{}}#2\end{tabular}}
  
\begin{document}

%
% The ''title'' command has an optional parameter, allowing the author to define a ''short title'' to be used in page headers.
\title{Spam Review Detection with Graph Convolutional Networks}

%
% The ''author'' command and its associated commands are used to define the authors and their affiliations.
% Of note is the shared affiliation of the first two authors, and the ''authornote'' and ''authornotemark'' commands
% used to denote shared contribution to the research.
% \author{Ben Trovato}
% \authornote{Both authors contributed equally to this research.}
% \email{trovato@corporation.com}
% \orcid{1234-5678-9012}
% \author{G.K.M. Tobin}
% \authornotemark[1]
% \email{webmaster@marysville-ohio.com}
% \affiliation{%
%   \institution{Institute for Clarity in Documentation}
%   \streetaddress{P.O. Box 1212}
%   \city{Dublin}
%   \state{Ohio}
%   \postcode{43017-6221}
% }

\author{Ao Li*}
% \authornote{Both authors contributed equally to this research.}
\affiliation{%
  \institution{Alibaba Group}
  \city{Hangzhou}
  \country{China}}
\email{jianzhen.la@alibaba-inc.com}

\author{Zhou Qin*}
% \authornote{Both authors contributed equally to this research.}
\affiliation{%
  \institution{Alibaba Group}
  \city{Hangzhou}
  \country{China}}
\email{qinzhou.qinzhou@alibaba-inc.com}

\thanks{*The first two authors contributed equally}

\author{Runshi Liu}
\affiliation{%
 \institution{Alibaba Group}
 \city{Hangzhou}
 \country{China}
}
\email{runshi.lrs@alibaba-inc.com}

% \author{Gaojie Cui}
% \affiliation{%
%   \institution{Alibaba Group}
%   \city{Hangzhou}
%   \country{China}}
% \email{gaojie.cgj@alibaba-inc.com}

\author{Yiqun Yang}
\affiliation{%
  \institution{Alibaba Group}
  \city{Beijing}
  \country{China}}
\email{yiqun.yyq@alibaba-inc.com}

\author{Dong Li}
\affiliation{%
 \institution{Alibaba Group}
 \city{Hangzhou}
 \country{China}
}
\email{shiping@taobao.com}

%
% By default, the full list of authors will be used in the page headers. Often, this list is too long, and will overlap
% other information printed in the page headers. This command allows the author to define a more concise list
% of authors' names for this purpose.
% \renewcommand{\shortauthors}{Trovato and Tobin, et al.}

%
% The abstract is a short summary of the work to be presented in the article.
\begin{abstract}
Customers make a lot of reviews on online shopping websites every day, e.g., Amazon and Taobao. Reviews affect the buying decisions of customers, meanwhile, attract lots of spammers aiming at misleading buyers. Xianyu, the largest second-hand goods app in China, suffering from spam reviews. The anti-spam system of Xianyu faces two major challenges: scalability of the data and adversarial actions taken by spammers. In this paper, we present our technical solutions to address these challenges. We propose a large-scale anti-spam method based on graph convolutional networks (GCN) for detecting spam advertisements at Xianyu, named GCN-based Anti-Spam (GAS) model. In this model, a heterogeneous graph and a homogeneous graph are integrated to capture the local context and global context of a comment. Offline experiments show that the proposed method is superior to our baseline model in which the information of reviews, features of users and items being reviewed are utilized. Furthermore, we deploy our system to process million-scale data daily at Xianyu. The online performance also demonstrates the effectiveness of the proposed method.

% To our knowledge, this is the first spam detection problem addressed by graph convolutional networks.

% \textit{\\\\ToDo need refine}

\end{abstract}

%
% The code below is generated by the tool at http://dl.acm.org/ccs.cfm.
% Please copy and paste the code instead of the example below.
%
% begin deleted
% \begin{CCSXML}
% <ccs2012>
% <concept>
% <concept_id>10010147.10010178</concept_id>
% <concept_desc>Computing methodologies~Artificial intelligence</concept_desc>
% <concept_significance>500</concept_significance>
% </concept>
% <concept>
% <concept_id>10010147.10010919</concept_id>
% <concept_desc>Computing methodologies~Distributed computing methodologies</concept_desc>
% <concept_significance>100</concept_significance>
% </concept>
% </ccs2012>
% \end{CCSXML}

% \ccsdesc[500]{Computing methodologies~Artificial intelligence}
% \ccsdesc[100]{Computing methodologies~Distributed computing methodologies}
% end deleted
%
% Keywords. The author(s) should pick words that accurately describe the work being
% presented. Separate the keywords with commas.
% \keywords{Spam detection, graph convolutional networks, heterogeneous graph}

%
% A ''teaser'' image appears between the author and affiliation information and the body 
% of the document, and typically spans the page. 
% \begin{teaserfigure}
%   \includegraphics[width=\textwidth]{sampleteaser}
%   \caption{Seattle Mariners at Spring Training, 2010.}
%   \Description{Enjoying the baseball game from the third-base seats. Ichiro Suzuki preparing to bat.}
%   \label{fig:teaser}
% \end{teaserfigure}

%
% This command processes the author and affiliation and title information and builds
% the first part of the formatted document.
\maketitle

\section{Introduction}

%三段论：
% 1. 提出问题说明重要性
% 2. 说明问题中存在的挑战
% 3. 说明自己的方法， 
%    a. 如何解决这些挑战 
%    b. 贡献点

%第一段：提出问题并说明重要性
Reviews of online shopping websites provide valuable information, such as product quality and aftersales service. These reviews, which straightforwardly influence purchase decisions of customers \cite{samha2014aspect}, have become a target place for spammers to publish malicious information. In our case, Xianyu, the largest second-hand goods app in China, which facilitates daily sales of over 200,000 products and achieves an annual Gross Merchandise Volume (GMV) over 13 billion dollars from August 2017 to July 2018, is also suffering from spam reviews. These spam reviews need to be cleaned out because they not only undermine experience of users but also provide a hotbed for internet fraud.

%第二段：说明问题的特殊性和挑战
%说明闲鱼上的review的不同 %列举闲鱼上的两种主要风险 指出我们要针对的风险。
Reviews at Xianyu differ from reviews in other e-commerce websites in several aspects. For instance, at Amazon or Taobao, reviews are usually made by customers who have bought the products, therefore review action usually happens \textit{after} purchase. In contrast, users have no idea about the quality and possible lowest price of the second-hand goods. Thus reviews at Xianyu act as a communication tool for buyers and sellers (e.g., query for details and discounts) and review action usually happens \textit{before} purchase, as shown in Figure \ref{fig:xianyu_screen} and Figure \ref{fig:abs_xianyu}. So instead of \textit{review} the term \textit{comment} will be used in the rest of paper to underline the essential difference of spam types at Xianyu. Generally, there are two main kinds of spam comments at Xianyu: \textit{vulgar comments} and \textit{spam advertisements}. Given the fact that spam advertisements takes the majority of spam comments, we focus on spam advertisements detection in this work,. 
% , which commonly mislead people to risky offline activities. In this work, we focus on spam advertisements detection problem. 

%指出我们要针对的问题中，挑战在哪里
%挑战有两个：1.数据规模 2.变异
The main challenges of spam advertisements detection are:
\begin{itemize}
    \item \textbf{Scalability}: Large-scale data of Xianyu with over 1 billion second-hand goods published by over 10 millions users. 
    \item \textbf{Adversarial Actions}: Same as most risk control system, the anti-spam system suffers from performance decay according to adversarial actions taken by spammers.
\end{itemize}
Spammers normally take the following two adversarial tricks to circumvent the anti-spam system: 
\begin{itemize}
    \item \textbf{Camouflage}:Using different expressions with similar meaning. For example, ``Dial this number for a part-time job'' and ``Want to earn more money in your spare time? Contact me'' are both spam advertisements for the purpose of guiding people to risky offline activities.
    \item \textbf{Deforming the comments}: Spammers replace some keywords in the comments with rarely used Chinese characters or typos deliberately. For example, ``Add my vx", ``Add my v" and ``Add my wx" all mean "Add my WeChat$\footnote{WeChat is the largest messaging and social media app in China and one of the world's largest standalone mobile apps by monthly active users.
    }$ account".

\begin{figure}[h]
    \centering
    \includegraphics[width=0.45\textwidth]{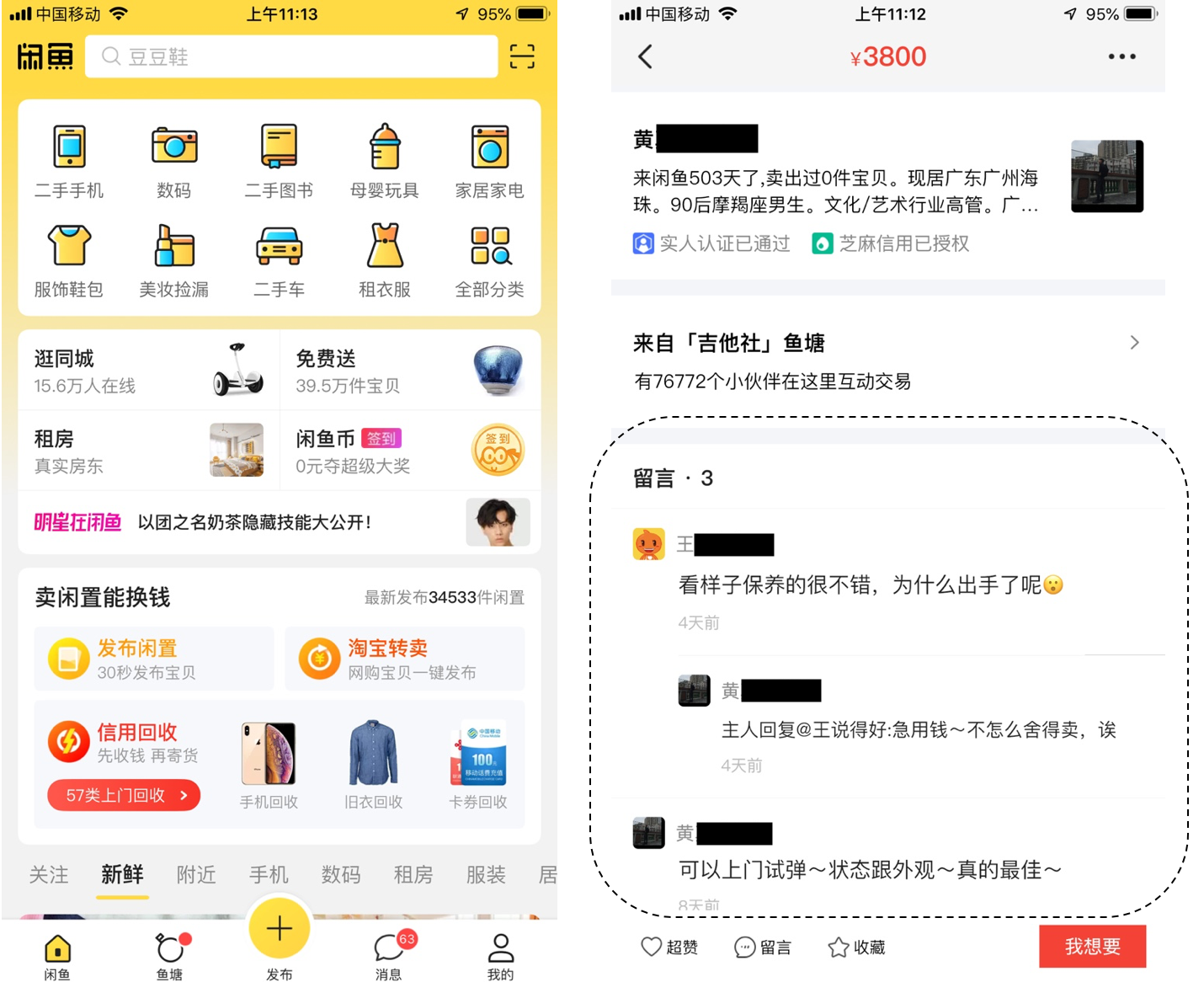}
    
    \caption{Xianyu App: the main page (left) and the comment area (right). The comment area highlighted with dashed rectangle provides the main communication tool of Xianyu.}
    \label{fig:xianyu_screen}
\end{figure}

\end{itemize} 
These tricks bring some inconvenience but still understandable for human readers. On the contrary, a big challenge for the anti-spam system is to recognize various patterns designed by spammers.

% 3. 说明自己的方法， 
At the same time, it's noticed that the impact of adversarial actions can be alleviated by introducing the context of the comments. We define the context into two categories: local context and global context. The local context refers to the information from the publisher and the item related, while the global context refers to the information offered by the feature distribution of all comments. 

In this work, we present a highly-scalable anti-spam method based on graph convolutional networks (GCN), dubbed GCN-based Anti-Spam method (GAS).

% we proposed and deployed that outperforms the baseline framework and alleviates the impact of adversary by introducing local and global context at Xianyu.

% Here we present a highly-scalable heterogeneous GCN framework we deployed at Alibaba, it is used to process a graph of over 200 million edges and 20 million nodes per day, and detect over 10000 spam reviews 
%\\\textit{TODO}\\
% (根据业务修改)
In summary, the contributions of the work are listed below:

\begin{enumerate}
    \item We propose a GCN-based heterogeneous graph spam detection algorithm which works on a bipartite graph with edge attributes at Xianyu. The designed model significantly outperforms the baseline model and can be easily generalized to a meta-path\cite{sun2012mining} based heterogeneous GCN algorithm for various heterogeneous graphs and applications. 

    \item Besides the heterogeneous graph which utilizes local context of comments, we make use of global context and propose GAS, which further improves the result.
    
    \item We deploy the proposed anti-spam model with distributed Tensorflow framework to process million-scale comments daily at Xianyu. According to offline experiments and online evaluation, our system remarkably identifies much more spam comments and alleviates the impact of adversarial actions while satisfying the efficiency requirement.
    
\end{enumerate}

The rest of the paper is organized as follows. Section \ref{sec:related} lists the related work. In Section \ref{sec:proposed}, we elaborate the proposed GAS model. Offline and online experiments are presented in Section \ref{sec:experiments}. We introduce the implementation and deployment of the system at Xianyu in Section \ref{sec:system}. The work is summarized in Section \ref{sec:conclusion}.

\section{Related Work}\label{sec:related}
% \subsection{Traditional methods for spam detection}
% 无关图的注释掉
% The first concept of ''review spam'' was proposed by Jindal et  al. \cite{jindal2008opinion}. They defined three types of review spam, namely, untruthful opinions, reviews on brands only, and non-reviews. Based on these three categorizations, the authors detected whether it is a spam review or not. After that, Many features-based methods have been proposed, in which authors use hand-designed features for further machine learning techniques including supervised, unsupervised and hybrid ways. 
% 讲spammer detection的 注释掉
% The detection of review spammer started by Lim et al. \cite{lim2010detecting}, they proposed a scoring technique method to find the ''spamicity'' degree of review spammer by predefining several abnormalities indicators (statistic features) extracted from review content and customers' ratings, then logistic regression was used to classify spam reviewers. The similar approach proposed by Huang et al. \cite{huang2013detecting} utilized review posting frequency (behavior feature) and reviews sentiment strength (linguistic feature) to construct two Spammer Detectors, then linearly combined them as the final detector. Nonlinear methods such as KNN also had been used, Xu et al. \cite{xu2013uncovering} make use of KNN and Markov network to detect collusive spammers in Chinese review website.

Most existing spam detection methods focus on extracting robust engineered features from review contents or reviewer behaviors. \cite{jindal2008opinion} studied the duplication of review content to detect spam reviews. They collected review centric, reviewer centric and product centric features, and fed them to a logistic regression model. \cite{ott2011finding} focused merely on the content of a review.  The authors approached the problem using three strategies as features in Naive Bayes and SVM classifier. \cite{liu2012survey} summarized domain expert features for opinion mining, then a set of elaborate designed features are used for review classification task.
These feature-centric methods ignore relations between reviewers, goods and comments. However, based on our observations, relations also play an important role in spam detection. For example, spam advertisements are often published by spammers in groups.

Based on similar observations, some scholars began to utilize the graph information. The first graph-based spam detection method was presented in \cite{wang2012identify}. They constructed the ``review graph'' with three types of nodes --- reviewers, stores, and reviews. Then reinforced the reviewer trustiness, store reliability and honesty of review in a HITS\cite{kleinberg1999hubs} like way. Liang et al.\cite{luckner2014stable}  made use of two graphs: one is the heterogeneous graph mentioned above, the other one represents the supportive or conflict relationship between the reviewers.
Soliman\cite{soliman2016adaptive} proposed a novel graph-based technique that detects spam using graph clustering on a constructed user similarity graph which encodes user behavioral patterns within its topology. The NetSpam framework\cite{shehnepoor2017netspam} defined different meta-path types on review graph and used them in classification. 

%两种图方法年代久远，删了
% Given the comment graph,  probabilistic graphical models had also been proposed and rapidly became one of the most popular methods among all graph-based method.
% Fei et al. \cite{fei2013exploiting} introduce Markov random field method, in which hidden nodes represent the ground truth to be identified with three stage --- non-spammer, mixed and spammer. Observed node shows the current state of data.
% Lu et al. \cite{lu2013simultaneously} present a review factor graph model to consolidate all the features to describe each review and reviewer then creating belief propagation between reviews and reviewers. 

% \subsection{Neural network methods on graph}
Recent years have witnessed a growing interest in developing deep-learning based algorithms on graph, including unsupervised methods\cite{perozzi2014deepwalk,grover2016node2vec,liao2018attributed} and supervised methods\cite{kipf2016semi,hamilton2017inductive,velickovic2018graph,li2018deeper}. One of the most prominent progress is known as GCN\cite{kipf2016semi}, in which features of nodes are aggregated from local neighbors. The ``graph convolution'' operator is defined as feature aggregation of one-hop neighbors. Through iterative convolutions, information propagates multiple hops away in the graph. GCN achieves significant improvements compared to previous graph-mining methods such as DeepWalk\cite{perozzi2014deepwalk}. 
After that, a great number of researchers have engaged in this area.
William et al.\cite{hamilton2017inductive} proposed GraphSAGE, an inductive framework that leverages node sampling and feature aggregation techniques to efficiently generate node embeddings for unseen data, which breaks the limitation of applying GCN in transductive settings.
Graph Attention Networks (GAT)\cite{velickovic2018graph} incorporates attention mechanism into GCN. By calculating attention coefficients among nodes, GAT allows each node to focus on the most relevant neighbors to make decisions. 

Most of the graph methods focus on the homogeneous graph, while in many real-world applications, data can be naturally represented as heterogeneous graphs. Heterogeneous graphs were seldom studied before and attract growing interests nowadays. EAGCN\cite{shang2018edge} calculates heterogeneous node embeddings using attention mechanism. This model focuses on the case where multiple types of links connecting nodes in a graph. The author proposed to use ``multi-attention'' --- each attention function considers neighbors defined only by a particular link type. Similarly, GEM\cite{liu2018heterogeneous} focuses on the case where there are multiple types of nodes. The author proposed an attention mechanism to learn the importance of different types of nodes. Specifically, they divided the graph into subgraphs by node types and calculated the contribution of each subgraph to the whole system as attention coefficients.

Graph methods have been applied in many domains, e.g., recommendation system\cite{grbovic2018real,ying2018graph,wang2018billion,zhao2017meta}, chemical properties prediction\cite{shang2018edge}, healthcare\cite{choi2017gram}, malicious accounts detection\cite{liu2018heterogeneous} and so on. In this paper, a GCN-based method is first applied to the spam review detection problem, to the best of our knowledge.

\section{Proposed Method}\label{sec:proposed}
% In this section, we first xx.

In this section, we first present preliminary contents of graph convolutional networks, then we illustrate the anti-spam problem at Xianyu. Finally, we will demonstrate our GAS method in two aspects: we first introduce how to extend GCN algorithm for heterogeneous graph and then illustrate GAS by further incorporating global context.

\subsection{Preliminaries}\label{sec:preliminaries}

% 介绍GAT

Previous work\cite{hamilton2017inductive,kipf2016semi,velickovic2018graph} focus mainly on homogeneous graphs. Let $\mathcal{G} = (\mathcal{V}, \mathcal{E})$ be a homogeneous graph with node $v \in \mathcal{V}$, edge $(v,v') \in \mathcal{E}$, node feature $x_v = h_v^0 \in \mathbb{R}^{d_0}$ for $v \in \mathcal{V}$ where $d_0$ denotes the feature dimension of the node. The hidden state of node ${v}$ learned by the $l$-th layer of the model is denoted as $h_v^l \in \mathbb{R}^{d_l}$, $d_l$ denotes the dimension of the hidden state at $l$-th layer. 

The GCN-based methods follow a layer-wise propagation manner. In each propagation layer, all the nodes update simultaneously. As summarized in \cite{xu2018representation, xu2018powerful}, a propagation layer can be separated into two sub-layers: aggregation and combination. In general, for a GCN with $L$ layers, aggregation and combination sub-layers at $l$-th layer $(l=1,2,\cdots L)$ can be written as:
%这段抄JK有点多, 用上标表示层数
\begin{equation}\label{eq:general_aggregate_layer}
    h_{N(v)}^l = \sigma\left( W^l \cdot AGG( \{h_{v'}^{l-1}, \forall v' \in N(v)\} )\right)
\end{equation}
\begin{equation}\label{eq:general_combine_layer}
    h_v^l = COMBINE\left( h_{v}^{l-1},  h_{N(v)}^{l} \right)
\end{equation}
where ${N(v)}$ is a set of nodes adjacent to $v$,  $AGG$ is a function used for aggregate embeddings from neighbors of node $v$, this function can be customized by specific models, e.g., max-pooling, mean-pooling\cite{hamilton2017inductive} or attention based weighted summation\cite{velickovic2018graph}. $W^l$ is a trainable matrix shared among all nodes at layer $l$. $\sigma$ is a non-linear activation function, e.g., Relu. $h_{N(v)}^l$ denotes the aggregated feature of node $v$'s neighborhood at $l$-th layer. $COMBINE$ function is used to combine self embedding and the aggregated embeddings of neighbors, which is also a custom setup for different graph models, e.g., concatenation as in GraphSAGE\cite{hamilton2017inductive}.

In GCN\cite{kipf2016semi} and GAT\cite{velickovic2018graph}, there are no explicit combination sub-layers. Self information of $v$ is introduced  by replacing $N(v)$ with $\tilde{N}(v)$ in Eq.\eqref{eq:general_aggregate_layer}, where $\tilde{N}(v) = v \cup N(v)$. Hence $COMBINE$ step actually happens inside of $AGG$ step. 

% $\tilde{\mathcal{G}}$ adds a self-loop to every node $v \in \mathcal{V}$.

\subsection{Problem Setup}\label{sec:problem_setup}

% 介绍业务场景, 引出图
Our purpose is to identify spam comments at Xianyu, which can be formulated to an edge classification problem on a directed bipartite graph with attributed nodes and edges.

\begin{figure}[h]
    \centering
    \includegraphics[width=0.45\textwidth]{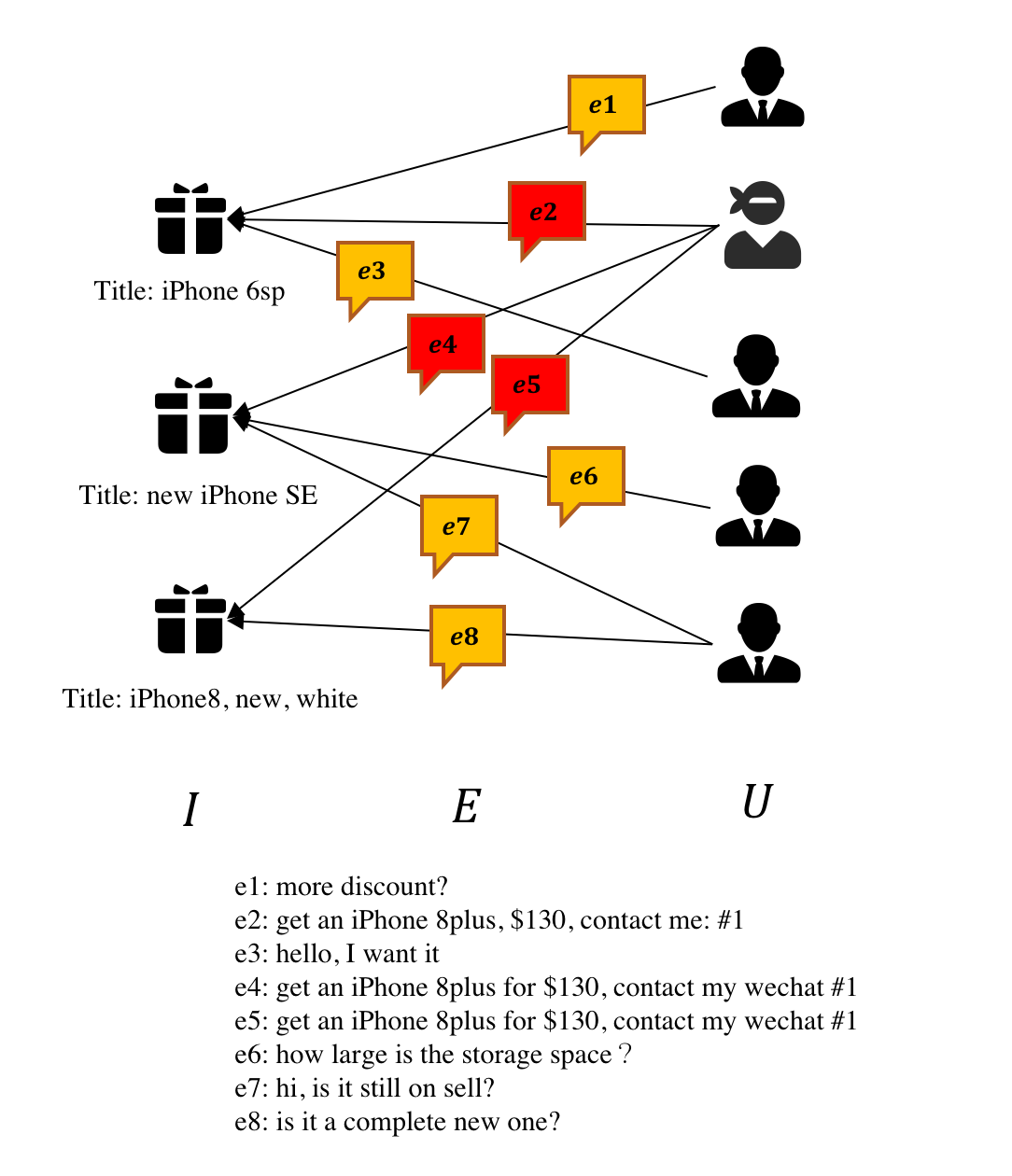}
    \caption{A miniature of Xianyu Graph. In this setting, spammer wants to mislead buyer to offline transactions, so he post an eye-catching comment saying that he has a cheaper phone for sale under many related items. $I, E, U$ represents item nodes, comments, user nodes respectively. Here \#1 stands for a specific WeChat account ID.
    % There are two types of comments, spam comments are marked with red background while non-spam comments are marked by green
    }
    \label{fig:abs_xianyu}
\end{figure}

% 说明图结构 引入 Notations, 参照【pixie】
Comments on Xianyu can be naturally represented as a bipartite graph. $G(U,I,E)$ where $U$ is the set of user nodes (vertices), $I$ is the set of item nodes (vertices) and $E$ is the set of comments (edges). An edge $e \in E$ from a user $u \in U$ to an item $i \in I$ exists if $u$ makes a comment $e$ to $i$. Additionally, given a vertex $v \in I \cup U$, let $N(v)$ be the set of vertices in node $v$'s one-hop neighbors, i.e. $N(v)=\{v'\in I\cup U|(v, v') \in E\}$. In the bipartite graph case of Xianyu, $N(i) \in U$ and $N(u) \in I$. $E(v)$ denotes the edges connected to $v$. Let $U(e)$ and $I(e)$ denote the user node and item node of edge $e$. This bipartite graph is named \textbf{Xianyu Graph}. See Figure \ref{fig:abs_xianyu} for a real word example.

%图结构示意图
% \begin{figure}[h]
%     \centering
%     \includegraphics[width=0.2\textwidth]{images/Xianyu.png}
%     % \input{figure/xianyu_graph.tex}
%     \caption{A miniature of Xianyu Graph. $I, E, U$ represents item nodes, comments, user nodes respectively. 
%     % There are two types of comments, spam comments are marked with red background while non-spam comments are marked by green
%     }
%     \label{fig:xianyu}
% \end{figure}

% At Xianyu, it is common that both spam and non-spam comments exist under the same item or published by the same user, so an aggregate function with attention mechanism is adopted to focus only on the related comments. 

% \subsection{GCN-based Anti-Spam Method}
\subsection{Heterogeneous Graph Convolutional Networks on Xianyu Graph} \label{sec:GAS}

As introduced in Section \ref{sec:preliminaries}, in the GCN-based node classification task on a homogeneous graph, node embedding from the last layer is used as the input of a classifier. 

\begin{figure}[h]
    \centering
    \includegraphics[width=0.48\textwidth]{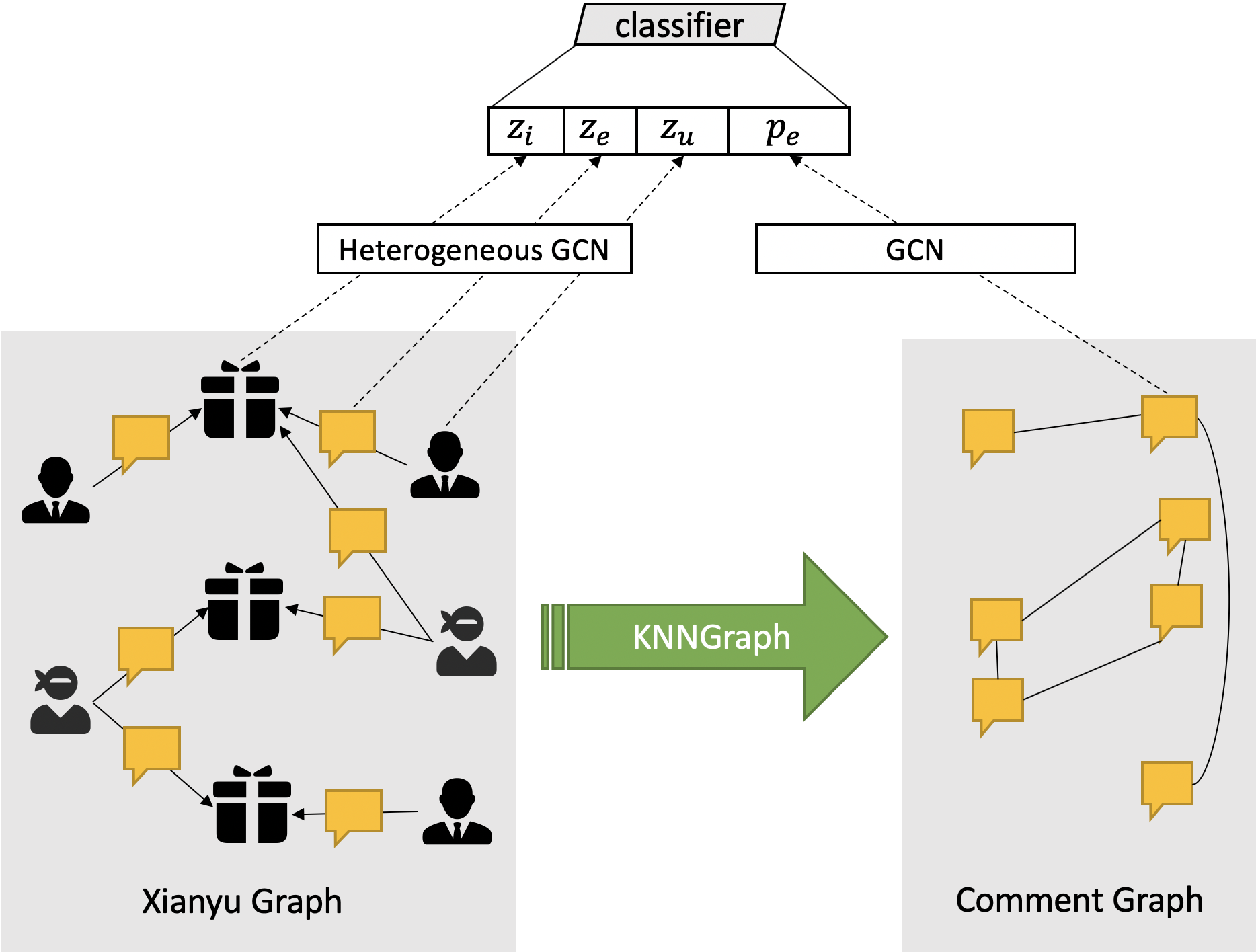}
    % \scalebox{0.8}{
    % \input{figure/two_graph.tex}
    % }
    \caption{An illustration of GAS which incorporates two graph models. Heterogeneous GCN which works on the Xianyu Graph provides user, item, comment embeddings $z_u, z_i, z_e$, respectively.  GCN which works on the homogeneous Comment Graph provides comment embedding $p_e$. In GAS, these embeddings are concatenated together as the input of the classifier: $y = classifier(concat(z_i, z_u, z_e, p_e))$.}
    \label{fig:two_graph_model}
\end{figure}

Instead, we utilize the edge embedding from the last propagation layer together with embeddings of the two nodes this edge links to. We concatenate these three embeddings for the edge classification task as shown in Figure \ref{fig:two_graph_model}, where $z_e$, $z_u$ and $z_i$  denote the edge, user and item embedding, i.e., $z_e = h_e^L$, $z_u = h_{U(e)}^L$ and $z_i = h_{I(i)}^L$.

We will concretely demonstrate how to tailor the standard GCN for a bipartite graph with edge attributes. The keypoint is to customize aggregation sub-layer and combination sub-layer in Eq.\eqref{eq:general_aggregate_layer} and Eq.\eqref{eq:general_combine_layer}.

\subsubsection{Aggregation Sub-layer}

Aggregation sub-layer in GCN treats all kinds of nodes the same and ignores the edge attributes. To fit the general framework Eq.\eqref{eq:general_aggregate_layer} to Xianyu Graph, three aggregation functions for each kind of entities (user, item, comment) are defined. 

For a comment, i.e., an edge, the hidden state is updated as the concatenation of previous hidden states of the edge itself and two nodes it links to. So the aggregation sub-layer is defined as 
\begin{equation}
\begin{split}\label{eq:heter_aggregation_edge}
    h_e^l = \sigma\left( W_E^l \cdot AGG^l_{E}( h_e^{l-1}, h_{U(e)}^{l-1}, h_{I(e)}^{l-1} )\right)
\end{split}
\end{equation}
where 
\begin{equation}
    AGG^l_{E}\left( h_e^{l-1}, h_{U(e)}^{l-1}, h_{I(e)}^{l-1} \right) = concat\left(h_e^{l-1}, h_{U(e)}^{l-1}, h_{I(e)}^{l-1}\right)
\end{equation}

For the user node $u \in U$ and item node $i \in I$, besides the information from neighbor nodes, the attributes of edges connected to them are also collected. The aggregated neighbor embedding $h_{N(u)}^l$ and $h_{N(i)}^l$ are calculated as
\begin{equation}
\begin{split}\label{eq:heter_aggregation}
    h_{N(u)}^l &= \sigma\left( W_U^l \cdot AGG^l_{U}( \mathcal{H}_{IE}^{l-1}  )\right) \\ 
    h_{N(i)}^l &= \sigma\left(  W_I^l \cdot AGG^l_{I}( \mathcal{H}_{UE}^{l-1} )\right)
\end{split}
\end{equation}
where
\begin{equation}
    \begin{split}
    \mathcal{H}_{IE}^{l-1}&=\left\{concat(h_i^{l-1}, h_e^{l-1}), \forall e=(u, i)\in E(u)\right\} \\
    \mathcal{H}_{UE}^{l-1}&=\left\{concat(h_u^{l-1}, h_e^{l-1}), \forall e=(u, i)\in E(i)\right\}
    \end{split}
\end{equation}
The two kinds of nodes maintain different parameters ($W_U^l, W_I^l$) and different aggregation functions ($AGG^l_{U}, AGG^l_{I}$).

As for the specific forms of $AGG^l_{U}$ and $AGG^l_{I}$, we adapt the attention mechanism:

\begin{equation}\label{eq:heter_attention}
\begin{split}
AGG^l_{U}( \mathcal{H}_{IE}^{l-1} ) & = ATTN_{U}\left(h_u^{l-1}, \mathcal{H}_{IE}^{l-1} \right) \\
    % & \forall i \in N(u), e \in E(u) \\
AGG^l_{I}( \mathcal{H}_{UE}^{l-1} ) & = ATTN_{I}\left(h_i^{l-1}, \mathcal{H}_{UE}^{l-1} \right) \\
    % & \forall u \in N(i), e \in E(i)
\end{split}
\end{equation}
$ATTN$ here is a function $f: h_{key}\times \mathcal{H}_{val} \rightarrow h_{val}$ which maps a feature vector $h_{key}$ and the set of candidates' feature vectors $\mathcal{H}_{val}$ to an weighted sum of elements in $\mathcal{H}_{val}$. The weights of the summation, i.e. attention values are calculated by the scaled dot-product attention \cite{vaswani2017attention}.

\subsubsection{Combination Sub-layer}

After aggregating the neighbors' information, we follow a combination strategy in \cite{hamilton2017inductive} for the user and item nodes as  
\begin{equation}\label{eq:heter_combination}
\begin{split}
    h_u^l &= concat\left(V_U^l\cdot h_u^{l-1}, h_{N(u)}^l\right) \\
    h_i^l &= concat\left(V_I^l\cdot h_i^{l-1}, h_{N(i)}^l\right)
\end{split}
\end{equation}
Where $V_U^l$ and $V_I^l$ denote trainable weight matrices for user node and item node, the $h_u^l$ and $h_i^l$ are the user hidden state and item hidden state of $l$-th layer. 
% Both of them will be fed into next layer together with $h_e^l$ in Eq.\eqref{eq:heter_aggregation}.

%jz
% The whole algorithm is described in Algorithm \ref{algo:heterogeneous_gnn}. Note that this method can actually be generalized to a meta-path based heterogeneous graph convolutional network algorithm with edge attributes, thus it can be easily applied to other applications abstracted by a heterogeneous graph with edge attributes. In our case and a 2-layer setting, two implicit meta-paths are $U\stackrel{E}{\longrightarrow}I\stackrel{E}{\longrightarrow}U$ and $I\stackrel{E}{\longrightarrow}U\stackrel{E}{\longrightarrow}I$. 

% jz expand
The whole algorithm is described in Algorithm \ref{algo:heterogeneous_gnn}. Note that this method can actually be generalized to  a meta-path based heterogeneous graph convolutional network algorithm for various heterogeneous graphs with edge attributes. In detail, for a meta-path $P$ in the form of $A^0\stackrel{R^0}{\longrightarrow}A^1\cdots A^{l-1}\stackrel{R^{l-1}}{\longrightarrow}A^{l}\stackrel{R^l}{\longrightarrow}\cdots\stackrel{R^{L-1}}{\longrightarrow}A^{L}$, where $A^l$ and $R^l$ corresponds to the node type and edge type on $P$. For a node $v$ of type $A^l$ at $l$-th layer on $P$, the aggregation and combination process can be written as:

% \begin{equation}\label{eq:heter_combination}
% \begin{split}
%     h_u^l &= concat\left(V_u^l\cdot h_u^{l-1}, h_{N(u)}^l\right) \\
%     h_i^l &= concat\left(V_i^l\cdot h_i^{l-1}, h_{N(i)}^l\right)
% \end{split}
% \end{equation}

\begin{equation}\label{eq:generalized_heter}
\begin{split}
    \mathcal{H}^{l-1}_{A^{l-1}E^{l-1}} &= \left\{concat\left(h^{l-1}_{v'}, h_e^{l-1}\right)  \quad \forall e \in (v, v')\in E_{R^{l-1}}(v) \right\} \\
    h^l_{N_{A^{l-1}}(v)} &= \sigma \left( W^l_{A^{l-1} \rightarrow A^l} \cdot AGG^l_{A^{l-1}\rightarrow A^{l}}\left( \mathcal{H}^{l-1}_{A^{l-1}E^{l-1}} \right) \right) \\
    h^l_{v} &= concat\left( V^l_{A^l} \cdot h^{l-1}_{v}, h^l_{N_{A^{l-1}}(v)} \right)
\end{split}
\end{equation}
where $E_{R^{l-1}}(v)$ denotes edges link to $v$ with edge type $R^{l-1}$ and $N_{A^{l-1}}(v)$ denotes neighbor nodes of $v$ with node type $A^{l-1}$. In our case and a 2-layer setting, two implicit meta-paths are $U\stackrel{E}{\longrightarrow}I\stackrel{E}{\longrightarrow}U$ and $I\stackrel{E}{\longrightarrow}U\stackrel{E}{\longrightarrow}I$.

\begin{algorithm}
\SetAlgoLined
% \dontprintsemicolon
\KwIn{Set of edges $E_b \subset E$, number of layers $L$, functions $U(E_b)$ and $I(E_b)$ which map $E_b$ to the user nodes and the item nodes $E_b$ linked, respectively. Xianyu Graph $G(U,I,E)$}
\KwOut{Hidden states of the $L$-th layer, include the hidden states of edges: $z_e, \forall e\in E_b$, the hidden states of users: $z_u, \forall u \in U(E_b) $ and the hidden states of items: $z_i, \forall i \in I(E_b)$}.
% states $z_e, \forall e\in E_b$, user states $z_u \forall u \in U(E_b) $ and items states $z_i \forall i \in I(E_b)$}
\Begin{ 
 $E^l \leftarrow E_b$ \;
 $U^l \leftarrow U(E_b)$ \;
 $I^l \leftarrow I(E_b)$ \;
 
 // Sampling \;
 %l代表层，0层需要更多的nodes
 \For{$l=L,\cdots,1$}{
    %自己赋值给底下一层
    $U^{l-1} \leftarrow U^{l}$ \; 
    $I^{l-1} \leftarrow I^{l}$ \; 
    
    %自己的邻居赋值给底下一层
    \For{$u \in U^l$}{ 
        $U^{l-1} \leftarrow U^{l-1} \cup SAMPLING( N(u))$ \;
    }
    
    %自己的邻居赋值给底下一层
    \For{$i \in I^l$}{ 
        $I^{l-1} \leftarrow I^{l-1} \cup SAMPLING( N(i))$ \;
    }
 }
 
 // Go through neural network \;
 %从1到L逐层进行forward
 \For{$l=1,\cdots,L$}{
    \For{$e \in E^l$}{
        $h_e^l = \sigma\left( W_E^l \cdot AGG^l_{E}( h_e^{l-1}, h_{U(e)}^{l-1}, h_{I(e)}^{l-1} )\right) $\;
    }
    
    \For{$u \in U^{l}$}{
        $\mathcal{H}_{IE}^{l-1} \leftarrow \left\{ concat(h_i^{l-1}, h_e^{l-1}), \forall e =(u,i) \in E(u) \right\}$ \;
        $h_{N(u)}^l \leftarrow \sigma\left( W_U^l \cdot AGG^l_{U}( \mathcal{H}_{IE}^{l-1}  )\right)$ \;
        $h_u^l = concat\left(V_U^l\cdot h_u^{l-1}, h_{N(u)}^l\right) $ \;
    }
    
    \For{$i \in I^{l}$}{
        $\mathcal{H}_{UE}^{l-1} \leftarrow \left\{ concat(h_u^{l-1}, h_e^{l-1}), \forall e =(u,i) \in E(i) \right\}$ \;
        $h_{N(i)}^l \leftarrow \sigma\left( W_I^l \cdot AGG^l_{I}( \mathcal{H}_{UE}^{l-1}  )\right)$ \;
        $h_i^l = concat\left(V_I^l\cdot h_i^{l-1}, h_{N(i)}^l\right) $ \;
    }
 }
 
 \For{$e \in E_b$}{
    $z_e = h_e^L$ \;
    $z_u = h_{U(e)}^L$ \;
    $z_i = h_{I(e)}^L$ \;
 }
 
 }%begin
 \caption{Heterogeneous Graph Convolutional Networks on Xianyu Graph.\label{algo:heterogeneous_gnn}}
\end{algorithm}

\begin{figure}[h]
    \centering
    \includegraphics[width=0.4\textwidth]{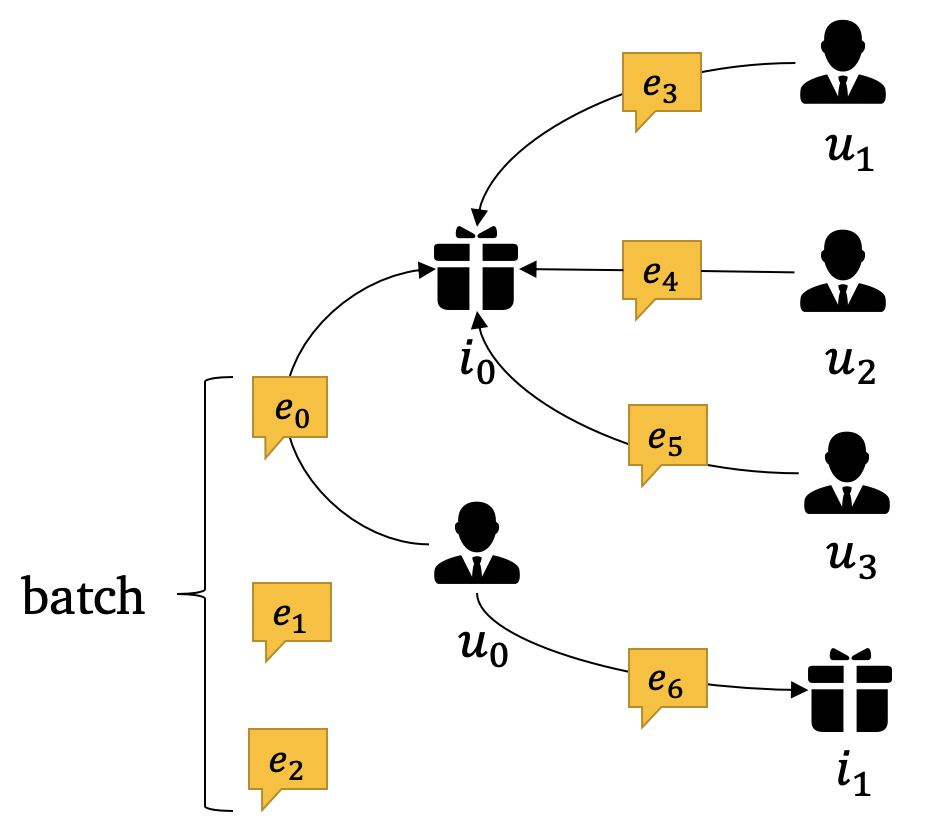}
    \caption{Time-related sampling strategy. Suppose $\{e_0, e_1, e_2\}$ form a batch of comments to be identified. To update the embedding of $e_0$, embeddings of item $i_0$ and user $u_0$ need to be calculated first. Without loss of generality, suppose we set the max number of samples $M=2$. For the item side, $2$ comments whose publish time are closest to the publish time of $e_0$ will be chosen from $\{e_3, e_4, e_5\}$, suppose $e_3, e_4$ are chosen, then $\{e_3, u_1\}$ and $\{e_4, u_2\}$ will be aggregated to $i_0$. Similarly, for the user side, e.g., $\{e_6, i_1\}$ with a padded placeholder will be chosen to aggregate to $u_0$.}
    \label{fig:sample}
\end{figure} 

\subsubsection{Time-related sampling strategy}
With the proposed aggregation sub-layer and combination sub-layer, either a whole-batch training strategy or a mini-batch training strategy can be conducted. The whole-batch training, which needs to update all the entities in one iteration, is impractical on massive data due to time consumption. Considering the scale of Xianyu Graph, the mini-batch training strategy is more suitable. For each item/user node, we sample a fixed number of neighbors to form a mini-batch feeding matrix as \cite{ying2018graph} does. Different from their random sampling strategy, we leverage time information and propose a time-related sampling as shown in Figure \ref{fig:sample}.

We summarize the sampling strategy in the following:
\begin{itemize}
    \item When the number of candidates is greater than the number of samples, i.e. $M$, we choose the closest $M$ comments in terms of time.
    \item When the number of candidates is less than $M$, we pad them with placeholders, and ignore all the computations related to these placeholders.
\end{itemize} 

Our sampling strategy is more reasonable than random sampling in two aspects. First, choosing closest comments is more reasonable than random subsampling since closest comments are more related to the comment to be identified. In the meanwhile, padding is more reasonable than resampling because comments posted under a second-hand good are often sparse. Padding avoids changing neighborhood distribution compared to resampling. In this way, we achieve a comparable result with a small $M$ thus saving training time as well as reducing memory consumption.

\subsubsection{Incorporate Graph Networks with Text Classification Model}

% 之前的工作中，节点和边的feature一般都是向量，而在我们的场景中，节点的feature是构造的特征向量，但是边的特征是文本，所以第一步是要将文本转化为向量
% https://github.com/brightmart/text_classification
% 有很多种方法可供选择【textcnn】【bert】，为了和图模型进行end-to-end的连接， 我们这里采用的是TextCNN, 一个效率高且效果好的模型。

% The spam detection can be intuitively regarded as a text classification problem with the additional user and item information. 
% 下面两个引文的引用逻辑有问题，引用应当是为了证明The processing of comments text is a typical text classification problem这句话，而且bert也不是纯classification任务
% The processing of comments text is a typical text classification problem\cite{bojanowski2017enriching}\cite{devlin2018bert}. 
% by extensive experiments 在学术论文中不要说
% By extensive experiments on different tasks, we found that

The text in comments should be converted into an embedding before being merged with user and item features. TextCNN\cite{kim2014convolutional} is a satisfactory text classification model that balance effectiveness and efficiency. Therefore we employ the TextCNN model to get comment embedding and integrate it to our graph neural network model as an end-to-end classification framework.

Specifically, we use word embedding pre-trained by word2vec\cite{mikolov2013distributed} as the input of TextCNN. The output of TextCNN is then used as the embedding of the comment. In detail, 
\begin{equation}
    h_e^0 = \textnormal{TextCNN}( w_0, w_1, w_2,\cdots ,w_n )
\end{equation}
where $w_i$ represents word embedding of $i$-th word of comment $e$, and $h_e^0$ is the initial embedding of comment $e$ in Eq.\eqref{eq:heter_aggregation_edge}. Therefore the parameters of TextCNN are trained together with others in the model described in Section \ref{sec:GAS}. 
% The process of text is shown in Figure \ref{fig:text2emb}.

% \begin{figure}[h]
%     \centering
%     % \includegraphics[scale=0.5]{images/text2emb.png}
%     \input{figure/text2emb.tex}
%     \caption{Process the text of comments.}
%     \label{fig:text2emb}
% \end{figure}

% \subsubsection{Incorporate two types of graph}

% In general, besides the heterogeneous Xianyu graph $G$ defined in Section \ref{sec:problem_setup}, we build the other homogeneous graph from $G$. Both of ${G}$ and ${G'}$ are utilized to provide local context and global context, respectively. 

% The outputs of the two algorithms are concatenate together before feeding into the standard softmax function. Specifically, given a comment $e$ to be classified, two nodes on $e$ can are $U(e)$ and $I(e)$. The final layer embeddings on $G$ is denoted as $z_e$,$z_{U(e)}$,$z_{I(e)}$ for $e$, $U(e)$, $I(e)$, respectively. For $G'$ instead, the embedding of final layer of $e$ is denoted as $p_e$, then the final embedding is 
% \begin{equation}
% 	y = classifier([\vec{Z}_i || \vec{Z}_u || \vec{Z}_e || \vec{P}_e]).
% \end{equation}
% See Figure \ref{fig:two_graph_model} for an illustration. 

% Incorporating Global Context and
\subsection{ GCN-based Anti-Spam model}\label{sec:EGAS}

% Incorporating Global Context and

% 问题是什么, local context失效 需要global
Spam comments at Xianyu are often deformed by malicious spammers as a countermove to our spam detection system. For example, deliberate typos and abbreviations are used to circumvent our detection. These spams have minor impact for human to read but often confuse our NLP model. Especially when spams are posted by different users and under different items. Obviously, the local context can not help in this situation. Intuitively, for this kind of spams, we want to find some supplementary information from the whole graph like ``How many similar comments on the whole site and what do they mean?". 

It's noticed that even if increasing the number of propagation layers help nodes capture the global context, the noise introduced can not be ignored, as other researchers reported\cite{kipf2016semi,li2018deeper}. Experiments in Section \ref{subsec:offline_eval} show that the performance can hardly benefit from increasing the number of propagation layers in our case. 

\begin{figure}[h]
    \centering
    \includegraphics[width=0.48\textwidth]{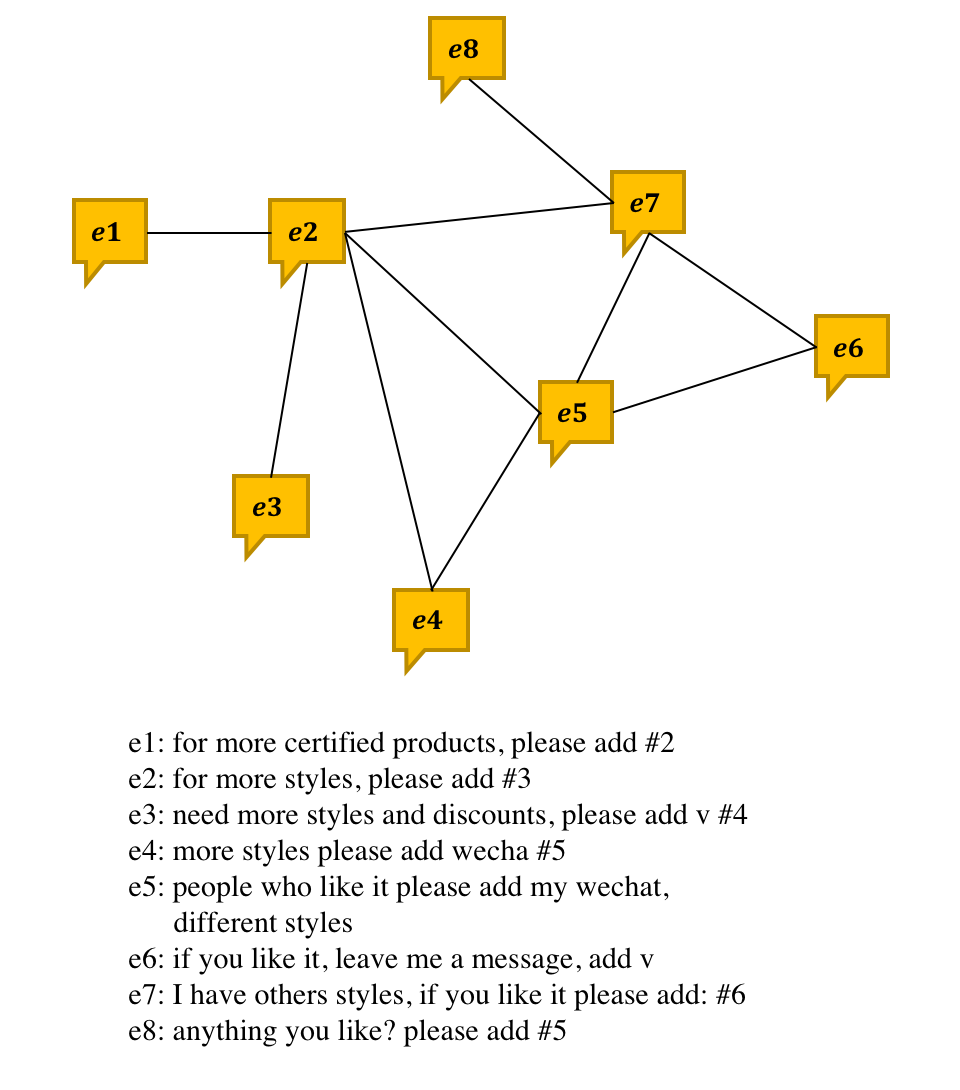}
    \caption{A miniature of Comment Graph, in which "wechat", "wecha" and "v" all mean WeChat account and \#2, \#3, \#4, \#5, \#6 stand for different account IDs, all the comments in this subgraph are spam advertisements.
    % There are two types of comments, spam comments are marked with red background while non-spam comments are marked by green
    }
    \label{fig:abs_comment_graph}
\end{figure}

Therefore, we takes a shortcut way to capture global context of nodes. More specifically, we construct a homogeneous graph named \textbf{Comment Graph} by connecting comments with similar contents. In this way, the edges(comments) in the heterogeneous Xianyu Graph now become vertices in Comment Graph. See Figure \ref{fig:abs_comment_graph} for a real world case of Comment Graph.

As demonstrated in \cite{li2018deeper}, GCNs on homogeneous graph can be viewed as a special form of Laplacian smoothing. Node classification tasks can benefit from this theory based on the assumption that nodes with same label are often grouped together in the graph, the features of nodes can be smoothed by its neighbors, thus make the classification task easier. Therefore an inductive GCN algorithm\cite{hamilton2017inductive} is performed on the constructed homogeneous Comment Graph to learn the comment embedding. 

By incorporating the inductive GCN which works on Comment Graph with the heterogeneous GCN which works on Xianyu Graph described in Section \ref{sec:GAS}, we propose an algorithm called GCN-based Anti-Spam(GAS) to train the model in an end-to-end manner. See Figure \ref{fig:two_graph_model} for the whole structure of GAS. 

Comment embedding of $e \in E$ learned by GCN from the Comment Graph is denoted as $p_e$. The final embedding of GAS is the concatenation of $p_e$ and other embeddings learned from Xianyu Graph, 
\begin{equation}
	y = classifier( concat(z_i, z_u, z_e, p_e) ),
\end{equation}
where  $z_e$, $z_u$ and $z_i$ denote the embeddings of $e$, $U(e)$ and $I(e)$ learned by the proposed heterogeneous GCN model, respectively. 

% See Figure \ref{fig:two_graph_model} for the whole structure of GAS. 

% 如何构建相似图
A non-trivial problem needed to be discussed is how to generate Comment Graph, namely, how to identify similar comments. This can be naively done by scanning all the comments, finding the closest k peers of each. However, it is impractical for the $\mathcal{O}(|E|^2)$ time complexity. In practice, we use approximate KNN Graph algorithm\cite{dong2011efficient} to construct a graph based on $K$ nearest neighbor of nodes.

In detail, the Comment Graph is constructed as follows:

\begin{itemize}
    \item Remove all the duplicated comments to avoid the trivial solution, i.e., two comments with same content are always most similar for each other.
    \item Generate comments embeddings by the method described in \cite{arora2017asimple}.
    \item Obtain the similar comment pairs by employing the approximate KNN Graph algorithm.
    \item Remove comment pairs posted by same user or posted under same item, since the local context has been taken into consideration on Xianyu Graph.
\end{itemize}

\begin{figure}[h]
    \centering
    \begin{subfigure}{.23\textwidth}
        \centering
        \includegraphics[width=\linewidth]{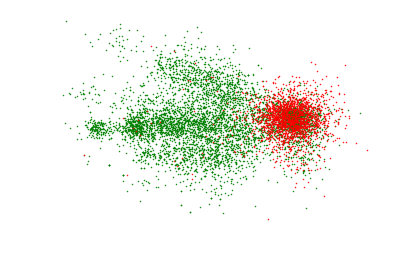}
        
        \label{fig:pca}
    \end{subfigure}
    \begin{subfigure}{.23\textwidth}
        \centering
        \includegraphics[width=\linewidth]{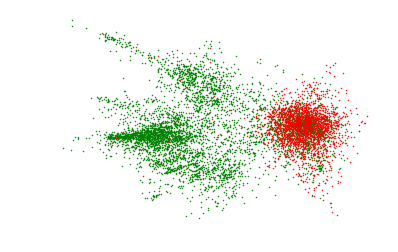}
        
        \label{fig:knngraph}
    \end{subfigure}
    \caption{Sample space visualization. Green points represent non-spam comments, and red points represent spam comments. Left: the original comment embeddings projected into 2-D space by PCA directly; Right: the smoothed comment embeddings (by averaging features of self and neighbors) projected into 2-D space by PCA.}
    \label{fig:comp}
\end{figure}

% 构建之后的效果
We assume in this way various spam reviews can be smoothed by integrating features of their neighbors. An visualization of a subset of training samples is provided to show that comments are more separable after the smooth process, see Figure \ref{fig:comp} for details.

% Intuitively, due to the adversarial actions of the spammers, the patterns of the spam comments differ from each other, that remind us to do that kind of smoothing by virtue of a graph where same labeled nodes are closer than nodes with different labels. 

% if we group all of these advertisements, then they actually form a complementary role with each other, make the identification much easier.

\begin{table}[h]
	
	\begin{tabular}{|l|c|c|}
		\hline
		Model feature& AUC & F1 score\\
		\hline
		Raw embedding& 0.9342& 0.8332\\
		Smoothed embedding&  0.9373& 0.8448\\
		\hline
	\end{tabular}
	\caption{Quantitative analysis of the effect of embedding smoothing on the subset of training samples. It clearly shows that samples with smoothed embedding are more separable for our task.}
	\label{tab:comp_quantity}
\end{table}

A quantitative analysis is also conducted to prove that comments are more separable after the smooth process. Two logistic regression models are trained and tested on raw embeddings and smoothed embeddings in Figure \ref{fig:comp}. AUC and F1-score are reported in Table \ref{tab:comp_quantity}. The results demonstrate that the linear separability of the samples is improved after smooth process. This improvement indicates that a classifier built on smoothed embedding will perform better.

%\subsubsection{Loss function}

\section{Experiments}\label{sec:experiments}

In this section, we evaluate the performance of our proposed method on true Xianyu dataset. First, we compare our method with several models using the offline dataset, then the online performance on Xianyu app is reported. Last, we present some real-world cases to give insight into the proposed methods.

\subsection{Offline Evaluation}\label{subsec:offline_eval}
\subsubsection{Dataset}

To evaluate the proposed method at Xianyu, we construct a heterogeneous graph with overall 37,323,039 comments published by 9,158,512 users on 25,107,228 items in a time period. We collect 1,725,438 regular comments along with 74,213 spam comments marked by human experts.

Training, validation and test set are randomly split with a ratio of 6:1:3. %splitting.

% \textit{TODO, how to sample the pos/neg samples}

\subsubsection{Comparing Methods}

% we call it Base-AS(Base AntiSpam model) to distinguish it from the proposed GAS model. 

% \begin{figure}[h]
%     \centering
%     % \includegraphics[scale=0.25]{images/baseline_model.png}
%     \input{figure/base_model.tex}
%     \caption{Baseline Model Structure.}
%     \label{fig:baseline_model}
% \end{figure}

To compare our method with traditional review mining method, we follow the instructions in \cite{liu2012survey}. Specifically, we design lots of hand-made features for comment, item and user(e.g. comment length, whether it is the first comment of the item, whether it is the only comment of the item, cosine similarity of the comment and item features, item price, number of comments made by the user, etc.). To encode the comment content, we calculate mutual information of each word. Top 200 words with largest mutual information values are selected out and then used to construct a binary value vector for each comment. Each entity of this vector indicate whether a word occurs or not. Lastly, we concatenate these features as the input of a GBDT model. We call this model GBDT as an abbreviation.

%jz
Instead of binary value encoding based on mutual information, a TextCNN model is also used to extract comment embedding. The comment embedding is then concatenated with user and item features described above as the input of a 2-layer MLP(Multilayer Perceptron) model. This was the model deployed online at Xianyu, thus it's regarded as the baseline model, named TextCNN+MLP.

% This is the model deployed at Xianyu was a language model TextCNN along with the features from user and item that directly related to this comment. 

To demonstrate the effectiveness of global context introduced by Comment Graph, we also compare the the model which only utilize the local context Xianyu Graph as in Section \ref{sec:GAS}. We call this model GAS-local.

In summary, the experiment configurations are detailed below:
% \begin{table}[h]
%     \centering
%     \begin{tabular}{|l|l|l|}
%         \hline
%          method& input features & classifier \\
%          \hline
%          Base-AS \textbf{*baseline} & TextCNN + raw user/item features    & 2-layer MLP\\
%          GBDT & MI + hand design user/item features & GBDT \\
%          GAS-local-1    &TextCNN + raw user/item features       &1 propagation layer on Xianyu Graph(i.e., 1-hop neighbors) \\
%          GAS-local-2 &TextCNN + user/item features       &2 propagation layers on Xianyu Graph. \\
%          GAS    &TextCNN + raw user/item features       &GAS model with 2 propagation layers on Xianyu Graph and 1 propagation layer on Comment Graph \\
%          \hline

%     \end{tabular}
%     \caption{detail co}
%     \label{tab:comp_config}
% \end{table}

\begin{itemize}
    \item GBDT: Domain expert features with GBDT model.
    \item TextCNN+MLP(\textbf{baseline}): TextCNN + user features + item features with 2 layer MLP model.
    \item GAS-local-1: 1 propagation layer on Xianyu Graph(i.e., 1-hop neighbors).
    \item GAS-local-2: 2 propagation layers on Xianyu Graph.
    \item GAS: GAS model with 2 propagation layers on Xianyu Graph and 1 propagation layer on Comment Graph.
\end{itemize}

The GBDT model is trained using 100 trees with a learning rate of 0.05. For other models, the TextCNN structure used by all the methods share the same hyperparameters, e.g., filter sizes are set to $\{3,4,5\}$ with 128 filters for each. All the methods except for GBDT are trained for 8 epochs. The max number of sampling $M$ is 16 for Xianyu Graph and the max number of sampling for Comment Graph is set to 64. The TextCNN+MLP model is trained with a stand-alone Tensorflow program and the learning rate is 0.001, while the proposed methods are all trained in a distributed manner with 8 GPUs. The learning rate is set to 0.005 and the batch size is set to 128.

\subsubsection{Result Analysis} \label{sec:result_analysis}

We evaluate these methods in terms of \textit{AUC}, \textit{F1-score} and \textit{recall at 90\% precision}. The metric \textit{recall rate at 90\% precision} is chosen since the detected spam reviews will be disposed in practice. We must ensure high model precision to avoid disturbing normal users. In our case, the precision over 90\% is an essential condition for a model to be deployed. so recall at 90\% precision becomes an essential criteria to compare different models.

\begin{table}[h]
    \centering
    \begin{tabular}{|l|c|c|c|}
        \hline
         method& AUC &F1 score &recall@90\% precision \\
         \hline
          GBDT &0.9649 &0.7686 & 50.55\%\\
         TextCNN+MLP  &0.9750       &0.7784  &54.86\%\\
         GAS-local-1                    &0.9806       &0.8138 &66.90\%\\
         GAS-local-2                     &0.9860       &0.8143 &67.02\%\\
         GAS                   &0.9872       &0.8217 &71.02\%\\
         \hline

    \end{tabular}
    \caption{Result comparison of offline experiments in terms of \textit{AUC}, \textit{F1-score} and \textit{recall at 90\% precision} which is denoted as \textit{recall@90\% precision}.} 
    \label{tab:result}
\end{table}

\begin{figure}[h]
    \centering
    \input{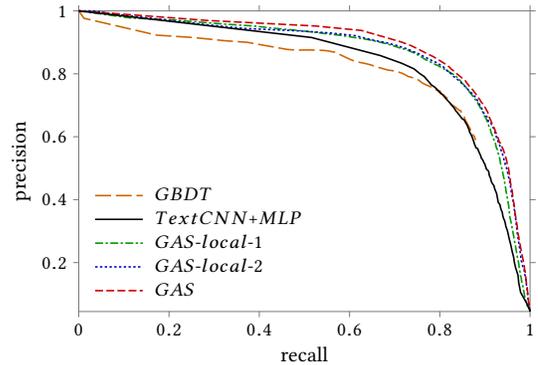}
    \caption{P-R chart of the offline evaluation. GAS, GAS-local-1 and GAS-local-2 significantly perform better and GAS gains further improvement compared to GAS-local-2.}
    \label{fig:prchart}
    % \centering
    % \input{figure/offline_zoom.tex}
    % \caption{P-R chart zoom out}
    % \label{fig:prchart_zoom}
\end{figure}

The results are shown in Table \ref{tab:result} and the PR curves are shown in Figure \ref{fig:prchart}. We can see that GAS-local-1, GAS-local-2 and GAS outperform the GBDT and TextCNN+MLP model in our dataset. This demonstrates the superiority of the proposed methods. The comparison of GAS-local-1 and TextCNN+MLP shows that the performance gain is significant attributed to the introduction of the local context. The $recall@90\% precision$ is improved from 54.86\% to 66.90\%, which means 12.04\% more spams can be detected and removed from the app. When comparing GAS-local-2 and GAS-local-1, the improvement is not prominent, which indicates the performance lift by incorporating 1-hop local context dominates. It may be due to the fact that besides the information introduced by 2-hop local context, noise is also introduced. As many authors had reported\cite{li2018deeper, kipf2016semi}, the performance gain attenuates dramatically as the hop increases. When comparing GAS and GAS-local-2, we can see a further improvement which demonstrates the effectiveness of incorporating of the global context. When precision is fixed to 90\%, compared to GAS-local-2, we detect extra 4\% spam comments. 

Overall, our proposed method outperforms the deployed baseline system with a 4.33 F1-score lift. Crucially, under the fixed disposal threshold of 90\% accuracy, our method brings an extra recall of 16.16\%.

\subsection{Online Performance}
We conduct online experiments on our platform as introduced in Section \ref{sec:system}. The goal of the experiment is to compare the number of spam comments detected by different models at 90\% precision which is checked by human experts. Detected spam comments will be clean out in production.

% while \textit{Base-AS} act as a real-time model and the other two ran once a day

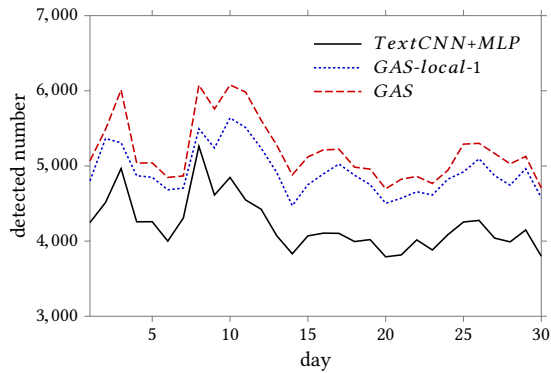
\begin{figure}[h]
    \centering
    \begin{tikzpicture}
\datavisualization [scientific axes,
x axis={label={day}, attribute=x, length=6cm},
y axis={label={detected number}, attribute=y, length=4cm, include value = {3000, 7000}},
style sheet=strong colors,
style sheet=vary dashing,
visualize as line/.list={a,c,b},
legend entry options/default label in legend path/.style=straight label in legend line,
legend={at values={x=22, y=6300}},
a={label in legend={text=$TextCNN$+$MLP$}},
b={label in legend={text=$GAS\textnormal{-}local\textnormal{-}1$}},
c={label in legend={text=$GAS$}},
]
data [set=a] {
	x, y
	1, 4247
	2, 4514
	3, 4965
	4, 4257
	5, 4258
	6, 4000
	7, 4307
	8, 5262
	9, 4613
	10, 4847
	11, 4548
	12, 4424
	13, 4072
	14, 3831
	15, 4069
	16, 4106
	17, 4103
	18, 3994
	19, 4020
	20, 3791
	21, 3816
	22, 4016
	23, 3882
	24, 4086
	25, 4254
	26, 4275
	27, 4040
	28, 3990
	29, 4148
	30, 3799
}
data [set=b] {
	x, y
	1, 4801
	2, 5365
	3, 5308
	4, 4870
	5, 4849
	6, 4683
	7, 4706
	8, 5494
	9, 5239
	10, 5639
	11, 5511
	12, 5236
	13, 4913
	14, 4471
	15, 4750
	16, 4894
	17, 5026
	18, 4877
	19, 4751
	20, 4503
	21, 4571
	22, 4657
	23, 4613
	24, 4824
	25, 4921
	26, 5095
	27, 4871
	28, 4744
	29, 4963
	30, 4579
}
data [set=c] {
	x, y
	1, 5068
	2, 5487
	3, 6011
	4, 5038
	5, 5041
	6, 4847
	7, 4867
	8, 6075
	9, 5758
	10, 6078
	11, 5983
	12, 5609
	13, 5277
	14, 4881
	15, 5120
	16, 5214
	17, 5222
	18, 4984
	19, 4958
	20, 4696
	21, 4823
	22, 4861
	23, 4767
	24, 4943
	25, 5290
	26, 5301
	27, 5167
	28, 5028
	29, 5127
	30, 4707
};

\end{tikzpicture}
    \caption{Online evaluation result. The proposed methods consistently outperforms the GBDT and TextCNN+MLP model at Xianyu.}
    \label{fig:online_performance}
\end{figure}

We deploy TextCNN+MLP, GAS-local-1 and GAS in our daily production environment and compare their performance. As depicted in Figure \ref{fig:online_performance}, the GAS-local-1 and GAS outperform TextCNN+MLP consistently in terms of the amount of detected spam comments. On the other hand, GAS outperforms GAS-local-1 consistently which further demonstrates that the system benefits from the global context introduced by the Comment Graph.

\subsection{Case Study}\label{sec:case_study}

The spam comments detected by different methods are manually checked.

% 加入Graph之后相对TextCNN多识别了什么
\subsubsection{GAS-local-1 vs. TextCNN+MLP}

We first compare the false negative samples of TextCNN+MLP (denoted as \textit{TextCNN+MLP}$_{FN}$) and the samples in \textit{TextCNN+MLP}$_{FN}$ recalled by GAS-local-1 (denoted as \textit{TextCNN+MLP}$_{FN} \cap$\textit{GAS-local-1}$_{TP}$. The result is shown in Table \ref{tab:graph_vs_cnn}.

\begin{table}[h]
    \centering
    \begin{tabular}{|l|c|}
        \hline
         samples & \specialcell{  average \#spams \\ within 1 hop on \\Xianyu Graph}   \\
         \hline
         \textit{TextCNN+MLP}$_{FN}$ & 2.60 \\
         
         \textit{TextCNN+MLP}$_{FN} \cap$\textit{GAS-local-1}$_{TP}$ & 3.24 \\
         \hline
    \end{tabular}
    \caption{Comparison of TextCNN+MLP and GAS-local-1. Obviously, the samples recalled by GAS-local-1 from \textit{TextCNN+MLP}$\bm{_{FN}}$ have more ``spam neighbors" in local context.} 
    \label{tab:graph_vs_cnn}
\end{table}

We analyze extra spam samples covered by GAS-local-1, finding that they are mostly similar advertisements published by the same people or under the same item. For instance, a typical spam comment is "check my profile photo for surprises", in which the spam information is hidden in the profile photo (The image information of profile photo is not used here because the profile photo is not contained in the comment and the time cost of image processing is high. Image information will be introduced in the future work). These advertisements alone contains no specific keywords and are not recognized by TextCNN+MLP. But GAS-local-1 correctly detects these advertisements by associating the comment with other comments published by this user.

% When a spammer tries to promote a product, he will repeatedly publish advertisements under many items to draw attention.

\subsubsection{GAS vs. GAS-local-1}

Likewise, we compare the false negative samples of GAS-local-1 (denoted as \textit{GAS-local-1}$_{FN}$) and the samples in \textit{GAS-local-1}$_{FN}$ recalled by GAS (denoted as \textit{GAS-local-1}$_{FN}\cap$\textit{GAS}$_{TP}$. The result is shown in Table \ref{tab:knn_vs_graph}.

\begin{table}[h]
    \centering
    \begin{tabular}{|l|c|c|}
        \hline
         samples & \specialcell{average \#spams\\within 1 hop on\\Xianyu Graph} & \specialcell{average \#spams\\within 1 hop on\\Comment Graph} \\
         \hline
         \textit{GAS-local-1}$_{FN}$ & 2.23 & 17.23 \\

         \textit{GAS-local-1}$_{FN}\cap$\textit{GAS}$_{TP}$ & 3.53 & 36.68\\
         \hline
    \end{tabular}
    \caption{Comparison of GAS and GAS-local-1. Obviously, the samples recalled by GAS from \textit{GAS-local-1}$\bm{_{FN}}$ have more ``spam neighbors" in global context.} 
    \label{tab:knn_vs_graph}
\end{table}

In detail, we analyze the results and find that two kinds of spam comments are more favoured by GAS compared to GAS-local-1:
\begin{itemize}
    \item \textbf{adversarial advertisements published by spammers}
    % This type of spammers are well grouped, they research the vulnerability of our anti-spam system exhaustively. 
    % Generally, they are no longer publish spam reviews through a single account. Alternatively, they collect many accounts then publish several advertisements for each account. 
    
    % What's more, they slightly change the syntax or words while keep meaning same. For example, a group of spammers try to ``Stock goods to deal with? contact me." or in another expression ``Contact me to deal with stock goods" or just ``clean up goods with a good price".  
    
    This kind of spam advertisements are deformed by spammers with similar meaning (see Table \ref{tab:knn_examples} for a typical example). As most of these spam reviews are not published by the same account or under the same item, but connected to each other on the Comment Graph. With the fixed disposal threshold of 90\% accuracy, they are not detected by GAS-local-1 but captured by GAS which takes advantage of the Comment Graph that introduces the global context. These spams may published by a group of spammers, they may collect many accounts and publish several advertisements using each account.

    \item \textbf{coupon messages published by different users} 
    
    This kind of coupon messages aim to lead people to another app. Once someone use the invitation code in the message, the publisher of the comment will be paid. Unlike the normal spammer that publishes a lot of similar comments under different items, this kind of spam comments is published by many different people, which may not be malicious spammers. But these coupon messages actually disturb other customers.
    This kind of publisher does not publish too much reviews as malicious spammers do, and will not gather under a particular item. That make it hard to recognize with only local context through Xianyu Graph. By introducing Comment Graph, this kind of similar comments group together, which would be recognized by GAS.
\end{itemize}

% case

% 32506764	教你撸神单：＋葳ling08882
% (2774733014,2743177562,2775371221,2740836391,2793094193,2771164915,2769934750)
% 

\begin{table}[h]
    \centering
    \begin{tabular}{|l|c|c|}
        \hline
         user\_id &item\_id & comment \\
         \hline
8737661	& 12381953 &	This is the bonus I got\\ 
& & at Taobao, contact me and I will\\
& & teach you how to do that \\
\hline
8737661	& 26771502 &	Get Bonus from Taobao, I can teach you \\
\hline
420310	& 27063522 &	Contact me to to learn to get \\
& & the bonus from Taobao \\
\hline
653613	& 20374180 &	Teach you to get bonus: +V xxxxxx \\
\hline
8806574	& 20634558 &	Vx:xxxxxx to teach you to get bonus \\
         \hline
    \end{tabular}
    \caption{Examples of extra spams detected by GAS compared to GAS-local, where ``+V xxxxxx" and ``Vx:xxxxxx" means WeChat app account.} 
    \label{tab:knn_examples}
\end{table}

%ent  加differ入KNN之后，多识别了什么

% 难以像图像模型等，可以展示具体的例子，这里只能给出多识别的样本是一个什么样的统计特征

\section{System Implementation And Deployment} \label{sec:system}

In this section, we introduce the implementation and deployment of the proposed GCN-based anti-spam method at Xianyu. We first give an overview of the system and then elaborate on the modules relevant to our method, especially the distributed Tensorflow model.

\subsection{System Overview}

\begin{figure}[h]
    \centering
    \includegraphics[width=0.45\textwidth]{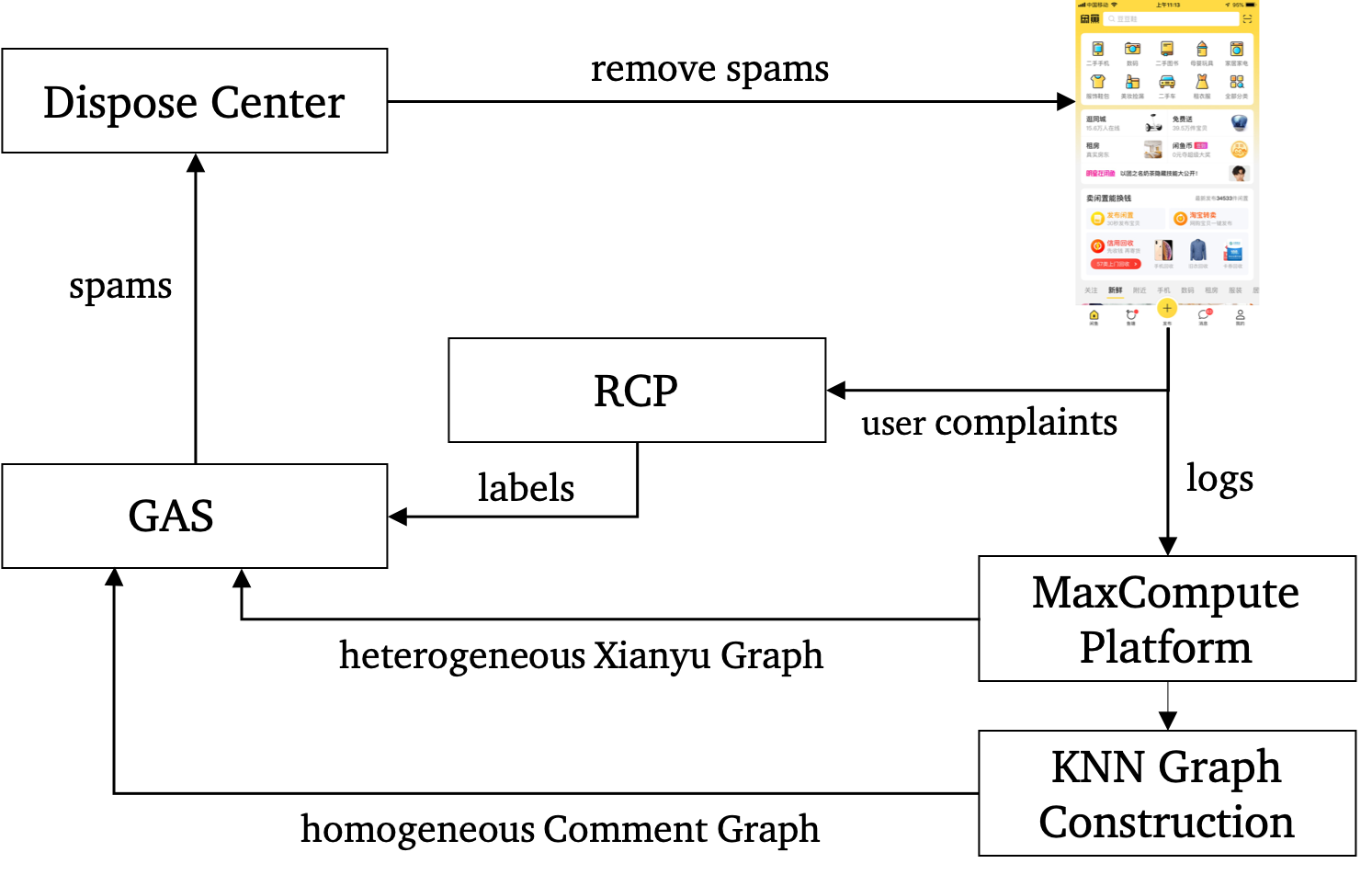}
    \caption{System Overview.}
    \label{fig:system}
\end{figure}

In Figure \ref{fig:system}, we show the architecture of the anti-spam platform at Xianyu. The workflow is illustrated in the following:
\begin{itemize}
    \item When a user comments on Xianyu App, the logs will be stored on the MaxCompute platform, which is a data processing platform for large-scale data warehousing. In practice, we choose the logs in the recent month to construct the heterogeneous graph per day.
    \item Based on the logs, the KNN Graph is constructed daily on the MaxCompute platform.
    \item A distributed Tensorflow implementation of GAS is employed to detect spams.
    \item The detected spams are removed from the app and the malicious accounts may be disabled.
    \item RCP (Risk Check Platform) is used to check the complaints from users being punished. The complaints reviewed and supported by human experts will lead to punishment withdrawal and be treated as mistakes of the model. The result of RCP will be used as labeled samples for further optimization of our model.
\end{itemize}

\subsection{Implementation}

% 总起

We conduct a distributed implementation of the proposed method. Considering the large scale data in Xianyu, i.e., billions of items, millions of users and billions of comments, the parameter-server architecture of Tensorflow is adopted to provide a distributed solution for storage, data-fetching, training and predicting. Specifically, we use 8 parameter servers along with 8 workers, each worker is equipped with an Nvidia V100 GPU card, 6 CPU cores, 32GB memory. Parameter server has 2 CPU cores with 300 GB memory each.

\subsubsection{storage}

% 如何存储

First, the graph data must be stored and readily available when the model needs to fetch the data from the graph. The Xianyu Graph is enormous and thus impractical to be saved in one machine, so the graph structure as well as the features of vertices and edges are saved in parameter servers. Note that graph structure is stored as adjacency list for memory efficiency.

% 分布式loading

A time-consuming step is data loading, which is both CPU-bound (parse) and I/O-bound (fetch). To accelerate the loading process, we split the adjacency list and feature matrices into several parts, evenly scattered them to parameter servers to perform a distributed loading task. Specifically, each worker is responsible for loading a particular part of the table to fill the corresponding sub-matrix in parameter server. By this means, a linear acceleration of $O(\#workers)$ is reached for loading data.

% We finally achieve a linear speedup on loading data this way.

% 好处

Scatter adjacency list and feature matrix evenly to parameter servers has another advantage: it helps balance the reading/writing load for the upcoming training process.

\subsubsection{data fetching}

% cache技术

During the computation period, workers will look up for necessary information on parameter server first, fetch them, then perform computations locally.  The parameter-server architecture avoids memory overflow while leading to less efficiency. Parameter server and workers are connected through the network but network throughput is far slower than the memory access. In our experiment, on each worker, there is about 41\% time wasted on fetching information from the server. To accelerate the lookup phase, we use cache mechanism on each worker, i.e., each worker will cache the features and adjacency lists locally when performing a search on parameter server. Cache technique save about 30\% training time.  

% \begin{figure}[h]
%     \centering
%     \includegraphics[scale=0.25]{images/ps.png}
%     \caption{Parameter Server}
%     \label{fig:ps}
% \end{figure}

% mini-batch获取技术

The adjacency lists and feature matrices are stored on parameter server. Even if we store a cache on each worker, frequently accessing neighborhood and feature information of nodes from CPU memory is not sufficient for GPU. So we follow the method introduced in \cite{ying2018graph}, which collects all the information that would be involved in the current mini-batch then feeding it to GPU memory at once. In this way, CPU-GPU communication during the computation is eliminated. Along with the producer-consumer mechanism employed in \cite{ying2018graph}, the GPU utilization is significantly improved.

% \subsubsection{computation} 

% The training and predicting process are also conducted in a distributed and synchronous optimization manner.

Finally, the training time of the offline experiment described in Section \ref{subsec:offline_eval} is reduced to 2 hours for GAS.

\section{CONCLUSION}\label{sec:conclusion}
The spam detection problem at Xianyu faces two main challeges: scalability and adversarial actions. To address these two challeges, we proposed an end-to-end GCN-based Anti Spam(GAS) algorithm which incorporates the local context and the global context of comments. The offline evaluation and online performance demonstrate the effectiveness of our method at Xianyu. Real-world cases are studied to further prove the effect of the different context introduced which alleviates the impact of adversarial actions. Finally, we elaborate on the implementation, deployment and workflow of the proposed method at Xianyu.

% The system is then deployed on Xianyu to support our daily anti-spam work. \textit{TODO, online evaluation}

\section{Acknowledgements}
We would like to thank Yuhong Li, Jun Zhu, Leishi Xu for their assistance on KNN Graph algorithm, and thanks to Huan Zhao for reviews and discussions.

%
% The next two lines define the bibliography style to be used, and the bibliography file.
\bibliographystyle{ACM-Reference-Format}
\bibliography{cikm}

\end{document}